\documentclass[twocolumn]{aastex631}

\usepackage{amsmath}
\usepackage{rotating,tabularx}
\defcitealias{Roberts-Borsani2022}{Paper I}
\defcitealias{Merlin2022}{Paper II}
\defcitealias{Castellano2022}{Paper III}
\defcitealias{Leethochawalit2022}{Paper X}

\shorttitle{The first rest-frame optical size-luminosity relation at $z>7$}
\shortauthors{Yang et al.}

\begin{document}

\title{Early results from GLASS-JWST. V: the first rest-frame optical size-luminosity relation of galaxies at $z>7$}

\correspondingauthor{Lilan Yang}
\email{lilan.yang@ipmu.jp}

\author[0000-0002-8434-880X]{L.~Yang}
\affiliation{Kavli Institute for the Physics and Mathematics of the Universe, The University of Tokyo, Kashiwa, Japan 277-8583}

\author[0000-0002-8512-1404]{T. Morishita}
\affiliation{Infrared Processing and Analysis Center, Caltech, 1200 E. California Blvd., Pasadena, CA 91125, USA}

\author[0000-0003-4570-3159]{N. Leethochawalit}
\affiliation{School of Physics, University of Melbourne, Parkville 3010, VIC, Australia}
\affiliation{ARC Centre of Excellence for All Sky Astrophysics in 3 Dimensions (ASTRO 3D), Australia}
\affiliation{National Astronomical Research Institute of Thailand (NARIT), Mae Rim, Chiang Mai, 50180, Thailand}

\author[0000-0001-9875-8263]{M.~Castellano}
\affiliation{INAF Osservatorio Astronomico di Roma, Via Frascati 33, 00078 Monteporzio Catone, Rome, Italy}

\author[0000-0003-2536-1614]{A. Calabr\`o}
\affiliation{INAF Osservatorio Astronomico di Roma, Via Frascati 33, 00078 Monteporzio Catone, Rome, Italy}

\author[0000-0002-8460-0390]{T. Treu}
\affiliation{Department of Physics and Astronomy, University of California, Los Angeles, 430 Portola Plaza, Los Angeles, CA 90095, USA}

\author{A.~Bonchi}
\affiliation{INAF Osservatorio Astronomico di Roma, Via Frascati 33, 00078 Monteporzio Catone, Rome, Italy}
\affiliation{ASI-Space Science Data Center,  Via del Politecnico, I-00133 Roma, Italy}

\author[0000-0003-3820-2823]{A. Fontana}
\affiliation{INAF Osservatorio Astronomico di Roma, Via Frascati 33, 00078 Monteporzio Catone, Rome, Italy}

\author[0000-0002-3407-1785]{C. Mason}
\affiliation{Cosmic Dawn Center (DAWN), Denmark}
\affiliation{Niels Bohr Institute, University of Copenhagen, Jagtvej 128, DK-2200 Copenhagen N, Denmark}

\author[0000-0001-6870-8900]{E.~Merlin}
\affiliation{INAF Osservatorio Astronomico di Roma, Via Frascati 33, 00078 Monteporzio Catone, Rome, Italy}

\author[0000-0002-7409-8114]{D.~Paris}
\affiliation{INAF Osservatorio Astronomico di Roma, Via Frascati 33, 00078 Monteporzio Catone, Rome, Italy}

\author[0000-0001-9391-305X]{M. Trenti}
\affiliation{School of Physics, University of Melbourne, Parkville 3010, VIC, Australia}
\affiliation{ARC Centre of Excellence for All Sky Astrophysics in 3 Dimensions (ASTRO 3D), Australia}

\author[0000-0002-4140-1367]{G. Roberts-Borsani}
\affiliation{Department of Physics and Astronomy, University of California, Los Angeles, 430 Portola Plaza, Los Angeles, CA 90095, USA}

\author[0000-0001-5984-0395]{M. Bradac}
\affiliation{University of Ljubljana, Department of Mathematics and Physics, Jadranska ulica 19, SI-1000 Ljubljana, Slovenia}
\affiliation{Department of Physics and Astronomy, University of California Davis, 1 Shields Avenue, Davis, CA 95616, USA}

\author[0000-0002-5057-135X]{E. Vanzella}
\affiliation{INAF - OAS, Osservatorio di Astrofisica e Scienza dello Spazio di Bologna, via Gobetti 93/3, I-40129 Bologna, Italy}

\author[0000-0003-0980-1499]{B. Vulcani}
\affiliation{INAF Osservatorio Astronomico di Padova, vicolo dell'Osservatorio 5, 35122 Padova, Italy}

\author[0000-0001-9002-3502]{D.~Marchesini}
\affiliation{
Department of Physics and Astronomy, Tufts University, 574 Boston Ave., Medford, MA 02155, USA}

\author[0000-0001-8917-2148]{X. Ding}
\affiliation{Kavli Institute for the Physics and Mathematics of the Universe, The University of Tokyo, Kashiwa, Japan 277-8583}

\author[0000-0003-2804-0648 ]{T. Nanayakkara}
\affiliation{Centre for Astrophysics and Supercomputing, Swinburne University of Technology, PO Box 218, Hawthorn, VIC 3122, Australia}

\author[0000-0003-3195-5507]{S. Birrer}
\affiliation{Kavli Institute for Particle Astrophysics and Cosmology and Department of Physics, Stanford University, Stanford, CA 94305, USA}
\affiliation{SLAC National Accelerator Laboratory, Menlo Park, CA, 94025}
\affiliation{Department of Physics and Astronomy, Stony Brook University, Stony Brook, NY 11794, USA}

\author[0000-0002-3254-9044]{K. Glazebrook}
\affiliation{Centre for Astrophysics and Supercomputing, Swinburne University of Technology, PO Box 218, Hawthorn, VIC 3122, Australia}

\author[0000-0001-5860-3419]{T. Jones}
\affiliation{Department of Physics and Astronomy, University of California Davis, 1 Shields Avenue, Davis, CA 95616, USA}

\author[0000-0003-4109-304X]{K.~Boyett}
\affiliation{School of Physics, University of Melbourne, Parkville 3010, VIC, Australia}
\affiliation{ARC Centre of Excellence for All Sky Astrophysics in 3 Dimensions (ASTRO 3D), Australia}

\author[0000-0002-9334-8705]{P. Santini}
\affiliation{INAF - Osservatorio Astronomico di Roma, via di Frascati 33, 00078 Monte Porzio Catone, Italy}

\author[0000-0002-6338-7295]{V. Strait}
\affiliation{Cosmic Dawn Center (DAWN), Denmark}
\affiliation{Niels Bohr Institute, University of Copenhagen, Jagtvej 128, DK-2200 Copenhagen N, Denmark}

\author[0000-0002-9373-3865]{X. Wang}
\affil{Infrared Processing and Analysis Center, Caltech, 1200 E. California Blvd., Pasadena, CA 91125, USA}

\begin{abstract}
We present the first rest-frame optical size-luminosity relation of galaxies at $z>7$, using the  NIRCam imaging data obtained by the GLASS James Webb Space Telescope Early Release Science (GLASS-JWST-ERS) program, providing the deepest extragalactic data of the ERS campaign.
Our sample consist of 19 photometrically selected bright galaxies with $m_\text{F444W}\leq27.8$ at $7<z<9$ and $m_\text{F444W}<28.2$ at $z\sim9-15$.
We measure the size of the galaxies in 5 bands, from the rest-frame optical ($\sim4800\,{\rm \AA}$) to the ultra-violet (UV; $\sim1600\,{\rm \AA}$) based on the S\'ersic model, and analyze the size-luminosity relation as a function of wavelength.
Remarkably, the data quality of NIRCam imaging is sufficient to probe the half-light radius $r_e$ down to $\sim 100$ pc at $z>7$.
Given the limited sample size and magnitude range, we first fix the slope to that observed for larger samples in rest-frame UV using HST samples.
The median size $r_0$ at the reference luminosity $M=-21$ decreases slightly from rest-frame optical ($600\pm80$ pc) to UV ($450\pm130$ pc).
We then re-fit the size-luminosity relation allowing the slope to vary. The slope is consistent with $\beta\sim0.2$ for all bands except F150W, where we find a marginally steeper slope of $\beta=0.53\pm0.15$. 
The steep UV slope is mainly driven by the smallest and faintest galaxies. If confirmed by larger samples, it implies that the UV size-luminosity relation breaks toward the faint end as suggested by lensing studies.

\end{abstract}

\keywords{galaxies: evolution – galaxies: fundamental parameters – galaxies: high-redshift – galaxies: structure }

\section{Introduction} \label{sec:intro}
The size of a galaxy is a fundamental observable quantity. Its evolution and scaling relation with other properties such as luminosity and stellar mass provide important clues on the formation and evolution of galaxies across cosmic time \citep{Conselice2014}.

At $z>7$, the study of the size-luminosity relation has been limited  so far to the rest-frame UV wavelength due to the lack of high-resolution imaging beyond the near-infrared bands. The nature of the size and size-luminosity relation in the rest-frame optical still remains an open question. 
Numerical studies predict a range of options for how the size of galaxies varies between UV and optical bands.
\citet{Ma2018} predicted that galaxy sizes measured in the rest-frame UV are smaller than those in optical because the UV emission tends to trace the clumpy star-forming regions, while \citet{Wu2020} predicted similar sizes in two bands.
In contrast, \citet{Marshall2022,Roper2022} found
that the perceived size of galaxies at the rest-frame UV 
is larger than in redder bands due to concentration of the dust in the central regions.
They predicted that the slope in the rest-frame optical band should be flatter than that in the UV band at these redshifts.

The UV sizes of $z>7$ galaxies (defined by half-light radius, $r_e$) are compact and marginally resolved by the Hubble Space Telescope (HST). 
Many studies have analyzed UV galaxy sizes and UV size-luminosity relation at $z>7$, expressed with parameterization $r_e\propto L^\beta$. Using HST data, \citet{Shibuya2015} found a constant $\beta=0.27\pm0.01$ across redshift range $0-8$,
which is consistent with many others \citep[][]{Grazian2012, Huang2013, Ono2013, Holwerda2015}, and a median size $r_0\sim420$ pc at $z\sim8$ at luminosity $M_\text{UV}=-21.0$.
the magnification power of gravitational lensing can boost the effective resolution of HST \citep{Kawamata2018,Bouwens2022,Yang2022}, but it cannot lift the limitation in wavelength coverage.

The James Webb Space Telescope (JWST) is revolutionizing our understanding of galaxies at $z>7$ by providing unprecedented depth and spatial resolution at both rest-frame optical and UV wavelengths.
The NIRCam filter F444W enables us to probe the rest-frame optical size of galaxies at $z>7$ at a similar angular resolution to that provided by HST WFC3-IR in the rest-frame UV i.e., F160W.
Based on observations from the GLASS James Webb Space Telescope Early Release Science (GLASS-JWST-ERS) program \citep{Treu2022}, we present the first measurement of the size of bright galaxies at $z>7$ across multiple broadband wavelengths. We derive their size-luminosity relation as a function of wavelength from rest-frame optical to UV. 

For this study, we limit ourselves to relatively bright/high signal-to-noise (SNR) galaxies, which alleviates the potential impact of incompleteness.
A more sophisticated analysis, including the full GLASS-JWST dataset (i.e., another half part of data will be achieved in the near future, see details in Section~\ref{sec:data}), a treatment of incompleteness, and evolutionary trends, is deferred to a forthcoming paper. A companion paper in this focus issue examines the wavelength dependency of morphological features such as clumpiness and asymmetry for the same sample \citep[][]{Treu2022b}.

We adopt a standard cosmology with $\Omega_{\rm m}=0.3$, $\Omega_{\Lambda}=0.7$ and H$_0$=70 km s$^{-1}$ Mpc$^{-1}$, and the AB magnitude system \citep{oke83,fukugita96}.

\section{JWST Program and Sample Selection} \label{sec:data}
\subsection{GLASS-JWST-ERS program}

The GLASS-JWST-ERS program is 
obtaining the deepest ERS observations with NIRISS, NIRSpec, and NIRCam \citep[][]{NIRISS,NIRSPEC1,NIRSPEC2,NIRCAM}.
In primary mode, it is obtaining NIRISS and NIRSpec spectroscopy 
of galaxies lensed by the HFF cluster Abell 2744 
in the cluster core field \citep[see details in Paper I,][]{Roberts-Borsani2022}.
In parallel mode, it is obtaining NIRCam imaging of galaxies 
in two flanking fields, i.e., 3-8 arcmins away from the center of the cluster, 
to NIRISS and NIRSpec, respectively.
We refer the reader to \citet{Treu2022} for details about the program and observational strategy.
GLASS-JWST-ERS utilizes 7 filters in NIRCam, 4 filters in the short wavelength channel (F090W, F115W, F150W, and F200W) and 3 filters (F277W, F356W and F444W) in the long wavelength channel.

In this letter, we focus on the NIRCam data obtained in parallel with  NIRISS on June 28-29 2022.
The data are described in more detail by \citet[][hereafter, Paper II]{Merlin2022}.
Modest lensing magnification is expected to present in the  parallel fields \citep{Medezinski2016,Bergamini2022}, so that the sizes and luminosities should be considered as upper limits. 
In this first analysis, we neglect this effect. The issue will be revisited after the completion of the campaign.

\subsection{Sample selection}

Our sample is composed of the combination of two photometrically-selected galaxies samples,
 of galaxies at $7<z<9$ \citep[][hereafter, Paper X]{Leethochawalit2022} and  of galaxies at $z\sim9-15$ \citep[][hereafter, Paper III]{Castellano2022}.
The samples are selected from
the photometric catalog of \citet[][hereafter, Paper II]{Merlin2022}.
We refer the reader to \citetalias{Merlin2022} for details of the image reduction and photometry.

Taking advantage of multiple filters, the catalogs employ the Lyman-break dropout technique and photometric redshift selection to make a robust selection of galaxies at $z>7$.
Both catalogs require $>8\sigma$ detection in F444W to ensure high purity of the selected candidates. The contamination rate of low-$z$ galaxies is expected to be $<10\%$. Based on the two catalogs, our sample consists of 13 galaxies at $7<z<9$ and 6 galaxies at $z\sim9-15$. Their $m_{\rm F444W}$ limits are $\leq 27.8$ and $<28.2$ respectively.

\section{Size Measurement}\label{sec:size}

To measure the size of each galaxy,
we utilize the \texttt{python} software \texttt{Galight}
\footnote{https://github.com/dartoon/galight} \citep{Ding2020}.
The code inherits the image modeling capabilities of \texttt{Lenstronomy} \citep{Birrer2015, Birrer&Amara2018, Birrer2021}, and adopts a forward modeling technique to measure the size of galaxies. 
The sizes derived by \texttt{Galight/Lenstronomy} are robust as demonstrated by \citet{Kawinwanichakij2021, Yang2021} and the results are consistent with those measured by traditional software such as \texttt{GALFIT} \citep{Peng2002}.

We model the light distribution of the sources assuming a S\'ersic light profile \citep{Sersic1968}:
\begin{equation}
I(r) = I_0 \exp \left[ -b_n \left(\frac{r}{r_e}\right)^{\frac{1}{n_{\text{s\'ersic}}}} - 1\right],
\end{equation} 
where $I_0$ the surface brightness amplitude at the half-light radius $r_\text{e}$, 
$b_n$ is a constant related to the S\'ersic index $n_\text{s\'ersic}$.
We adopt the definition, $r=\sqrt{x^2+y^2/q^2}$, where the coordinates $(x,y)$ depend on the position angle and the center position of the light, and $q$ is the axial ratio.
In addition, we stack several bright stars in the same field and use it as a point-spread function (PSF) in the forward modeling.

\begin{figure*} 
\centering
\includegraphics[trim={7cm 1cm 2cm 0cm},clip,width=2\columnwidth]{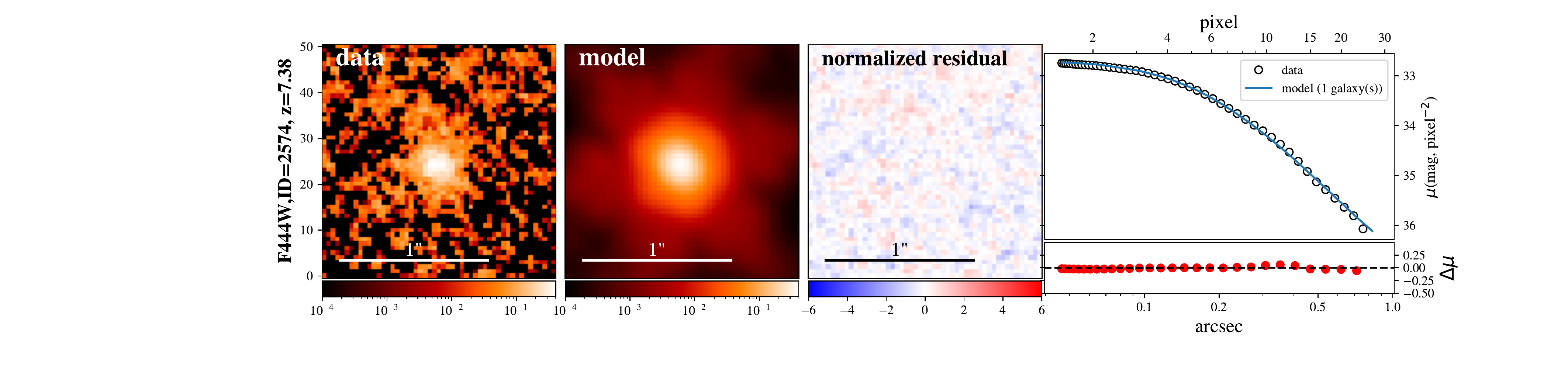}
\includegraphics[trim={7cm 1cm 2cm 0cm},clip,width=2\columnwidth]{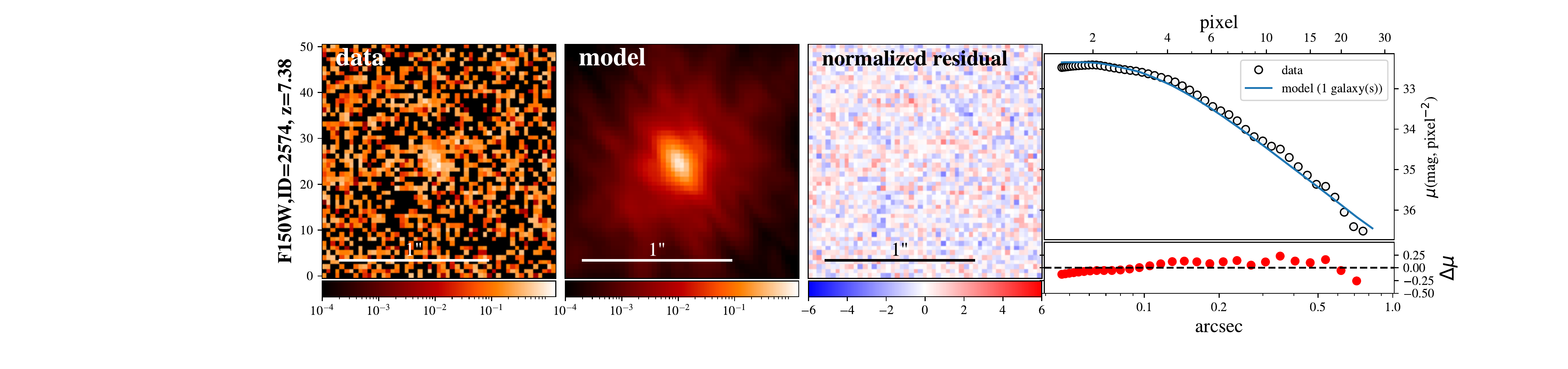}

\caption{Example of size measurements of a galaxy (ID=2574, $z=7.38$) in F444W (rest-frame optical; upper) and F150W (UV; bottom) bands using \texttt{Galight}. In each panel, from left to right, the columns represent (1) observed data, (2) best-fit model image, (3) normalized residual map, and (4) surface brightness profiles in 1D and its residual.}

\label{fig:size-fitting}
\end{figure*}

We use the mosaics images, with common aligned pixel grid (pixel size is 31 mas) in each band to measure the size of each galaxy in a $1.\!''6\times1.\!''6$ cutout,
see details of mosaics images in \citetalias{Merlin2022}.
We fix $n_\text{s\'ersic}$ to 1, which is appropriate for star-forming galaxies \citep{Morishita2014, Shibuya2015}, and the axial ratio is limited to the range 0.1-1.
In some cases, the cutout includes contamination, and we mask out or fit contaminants by additional S\'ersic light profiles.
We apply the same strategy to fit the galaxies in each band independently.
We visually inspect the fitting result for each galaxy in the sample in each band. We flag and remove the bands in which we are not confident of the size measurement owing to a poor fit or edge effects or defects.
Our results are summarized in Table~\ref{tab:results}.

We show an example of the modeling procedures in the rest-frame optical (F444W) and UV (F150W) bands in Figure~\ref{fig:size-fitting}.
In general, the galaxies are well fitted by an
exponential disk with $n_\text{s\'ersic}=1$.
We also perform the fitting while allowing $n_\text{s\'ersic}$ to vary in the range 0.3-4 and find there is no significant change in size on average ($<0.02$ dex).
More detailed morphology analysis can be found in a companion paper \citep{Treu2022b}.

\begin{sidewaystable*}
\centering
\scriptsize	
\caption{Size and luminosity of galaxies at $z>7$ measured from the rest-frame optical to the UV band. 
The columns represent the corresponding catalogs ID, coordinates, and photometric  redshift, provided in \citetalias[][]{Leethochawalit2022} for galaxies at $7<z<9$ and \citetalias[][]{Castellano2022} for galaxies at $z\sim9-15$, measured half-light radius $r_e$ and luminosity in the F444W, F356W, F277W, F200W and F150W, respectively, and $M_\text{UV}$.  }\label{tab:results}
\begin{tabular}{lcccccccccccccc}
\hline
\hline
ID & RA & Dec & $z_\text{ph}$ & $r_{e,\text{F444W}}$ & $M_\text{F444W}$ & $r_{e,\text{F356W}}$ & $M_\text{F356W}$ & $r_{e,\text{F277W}}$ & $M_\text{F277W}$ & $r_{e,\text{F200W}}$ & $M_\text{F200W}$ & $r_{e,\text{F150W}}$ & $M_\text{F150W}$ & $M_\text{UV}^{c}$\\
 &  &  &  & (kpc) &  &(kpc) &  & (kpc) & & (kpc) &  & (kpc)&  \\
\hline
1470 & 3.5137 & -30.3628 & 7.65 & 0.66 $\pm$ 0.09 & -19.0 $\pm$ 0.13 & 0.59 $\pm$ 0.11 & -18.9 $\pm$ 0.13 & 0.46 $\pm$ 0.12 & -18.2 $\pm$ 0.26 & 0.69 $\pm$ 0.20$^{a}$ & -18.6 $\pm$ 0.32 & \nodata & \nodata & -18.7 $\pm$ 0.25\\
2236 & 3.4899 & -30.3545 & 7.99 & 0.20 $\pm$ 0.04 & -18.8 $\pm$ 0.12 & 0.12 $\pm$ 0.05$^{a}$ & -17.9 $\pm$ 0.24 & 0.18 $\pm$ 0.07$^{a}$ & -18.4 $\pm$ 0.30 & 0.14 $\pm$ 0.04 & -18.5 $\pm$ 0.31 & 0.11 $\pm$ 0.03$^{a}$ & -18.6 $\pm$ 0.27 & -18.7 $\pm$ 0.20\\
2574 & 3.4955 & -30.3508 & 7.38 & 0.45 $\pm$ 0.06 & -19.3 $\pm$ 0.13 & 0.39 $\pm$ 0.06 & -19.3 $\pm$ 0.08 & 0.39 $\pm$ 0.08 & -19.0 $\pm$ 0.12 & 0.43 $\pm$ 0.07 & -19.4 $\pm$ 0.13 & 0.34 $\pm$ 0.06 & -18.9 $\pm$ 0.17 & -19.2 $\pm$ 0.15\\
2911 & 3.5118 & -30.3468 & 6.94 & 0.46 $\pm$ 0.02 & -20.7 $\pm$ 0.04 & 0.43 $\pm$ 0.02 & -20.7 $\pm$ 0.02 & 0.36 $\pm$ 0.03 & -20.1 $\pm$ 0.05 & 0.36 $\pm$ 0.05 & -19.9 $\pm$ 0.09 & 0.23 $\pm$ 0.03 & -19.5 $\pm$ 0.10 & -19.5 $\pm$ 0.15\\
2936 & 3.5119 & -30.3467 & 7.23 & 0.61 $\pm$ 0.04 & -20.0 $\pm$ 0.07 & 0.61 $\pm$ 0.04 & -20.0 $\pm$ 0.05 & 0.68 $\pm$ 0.06 & -19.8 $\pm$ 0.08 & 0.90 $\pm$ 0.13 & -19.7 $\pm$ 0.14 & 0.62 $\pm$ 0.11 & -19.5 $\pm$ 0.15 & -19.4 $\pm$ 0.30\\
3120 & 3.5203 & -30.3439 & 7.45 & 0.64 $\pm$ 0.06 & -19.7 $\pm$ 0.07 & 0.92 $\pm$ 0.11 & -19.9 $\pm$ 0.08 & 0.73 $\pm$ 0.08 & -19.8 $\pm$ 0.09 & 0.67 $\pm$ 0.08 & -19.7 $\pm$ 0.12 & 0.67 $\pm$ 0.11 & -19.5 $\pm$ 0.15 & -20.1 $\pm$ 0.20 \\
4542 & 3.4880 & -30.3254 & 8.97 & 0.47 $\pm$ 0.07 & -19.4 $\pm$ 0.12 & 0.17 $\pm$ 0.09 & -18.6 $\pm$ 0.19 & 0.40 $\pm$ 0.10$^{a}$ & -19.2 $\pm$ 0.13 & 0.14 $\pm$ 0.03 & -18.9 $\pm$ 0.18 & 0.10 $\pm$ 0.02 & -19.0 $\pm$ 0.17 & -19.5 $\pm$ 0.15\\
4863 & 3.4867 & -30.3272 & 8.07 & 0.24 $\pm$ 0.05 & -19.3 $\pm$ 0.12 & 0.35 $\pm$ 0.09 & -18.9 $\pm$ 0.12 & 0.20 $\pm$ 0.09 & -18.7 $\pm$ 0.14 & 0.27 $\pm$ 0.04 & -19.3 $\pm$ 0.12 & 0.18 $\pm$ 0.03 & -19.2 $\pm$ 0.15 & -19.3 $\pm$ 0.10\\
5001 & 3.4997 & -30.3177 & 8.10 & 1.01 $\pm$ 0.20 & -19.6 $\pm$ 0.13 & 1.40 $\pm$ 0.22 & -19.5 $\pm$ 0.18 & 3.14 $\pm$ 1.10$^{a}$ & -19.8 $\pm$ 0.23 & 1.07 $\pm$ 0.28 & -19.7 $\pm$ 0.20 & 1.23 $\pm$ 0.32$^{a}$ & -19.9 $\pm$ 0.18 & -19.4 $\pm$ 0.15\\
1708 & 3.4906 & -30.3604 & 7.83 & 0.52 $\pm$ 0.03 & -20.1 $\pm$ 0.09 & 1.04 $\pm$ 0.85$^{a}$ & -19.7 $\pm$ 0.47 & 0.41 $\pm$ 0.06 & -19.5 $\pm$ 0.09 & 0.32 $\pm$ 0.04 & -19.5 $\pm$ 0.12 & \nodata & \nodata & -19.5 $\pm$ 0.20\\
4397 & 3.4747 & -30.3226 & 8.06 & 0.87 $\pm$ 0.08 & -19.9 $\pm$ 0.14 & 0.34 $\pm$ 0.35$^{a}$ & -18.6 $\pm$ 0.79 & 0.65 $\pm$ 0.15$^{a}$ & -19.1 $\pm$ 0.18 & 1.11 $\pm$ 0.11 & -19.8 $\pm$ 0.17 & \nodata & \nodata & -19.4 $\pm$ 0.25\\
6116 & 3.5046 & -30.3079 & 8.24 & 0.32 $\pm$ 0.05 & -19.4 $\pm$ 0.09 & 0.72 $\pm$ 0.19$^{a}$ & -19.1 $\pm$ 0.16 & 4.34 $\pm$ 0.49$^{a}$ & -20.4 $\pm$ 0.24 & \nodata & \nodata  & 0.40 $\pm$ 0.28$^{a}$ & -18.4 $\pm$ 0.27 & -18.8 $\pm$ 0.30\\
6263 & 3.4697 & -30.3090 & 8.21 & 0.54 $\pm$ 0.07 & -19.5 $\pm$ 0.14 & 0.95 $\pm$ 0.76$^{a}$ & -19.5 $\pm$ 0.45 & 0.39 $\pm$ 0.12 & -19.1 $\pm$ 0.16 & 0.37 $\pm$ 0.10$^{a}$ & -19.2 $\pm$ 0.15 & \nodata & \nodata & -19.3 $\pm$ 1.90\\
560$^{b}$ & 3.5119 & -30.3718 & 10.63 & 0.50 $\pm$ 0.02 & -21.2 $\pm$ 0.05 & 0.46 $\pm$ 0.02 & -21.0 $\pm$ 0.09 & 0.49 $\pm$ 0.03 & -21.0 $\pm$ 0.11 & 0.43 $\pm$ 0.02 & -21.0 $\pm$ 0.15 & 0.37 $\pm$ 0.03 & -20.6 $\pm$ 0.20 & -21.0 $\pm$ 0.06\\
5153$^{b}$ & 3.4990 & -30.3248 & 12.30 & 0.17 $\pm$ 0.02 & -20.4 $\pm$ 0.04 & 0.10 $\pm$ 0.02 & -20.4 $\pm$ 0.06 & 0.12 $\pm$ 0.02 & -20.6 $\pm$ 0.07 & 0.12 $\pm$ 0.01 & -21.1 $\pm$ 0.10 & \nodata & \nodata & -21.2 $\pm$ 0.20\\
1351$^{b}$ & 3.5289 & -30.3638 & 9.33 & 0.76 $\pm$ 0.05 & -20.5 $\pm$ 0.07 & 0.78 $\pm$ 0.06 & -20.7 $\pm$ 0.08 & 0.67 $\pm$ 0.08 & -20.4 $\pm$ 0.13 & 0.88 $\pm$ 0.09 & -20.8 $\pm$ 0.13 & 0.68 $\pm$ 0.15 & -19.7 $\pm$ 0.44 & -20.7 $\pm$ 0.09\\
2488$^{b}$ & 3.5137 & -30.3516 & 9.93 & 2.01 $\pm$ 0.37$^{a}$ & -19.9 $\pm$ 0.12 & 1.07 $\pm$ 0.34 & -19.4 $\pm$ 0.28 & 0.39 $\pm$ 0.14$^{a}$ & -18.9 $\pm$ 0.32 & 0.39 $\pm$ 0.09 & -19.3 $\pm$ 0.34 & \nodata & \nodata & -20.0 $\pm$ 0.27\\
4131$^{b}$ & 3.4944 & -30.3076 & 9.20 & 0.28 $\pm$ 0.11$^{a}$ & -18.5 $\pm$ 0.08 & 0.17 $\pm$ 0.07$^{a}$ & -18.6 $\pm$ 0.09 & 0.19 $\pm$ 0.12$^{a}$ & -17.9 $\pm$ 0.11 & 0.21 $\pm$ 0.08$^{a}$ & -18.9 $\pm$ 0.19 & 0.16 $\pm$ 0.07$^{a}$ & -18.7 $\pm$ 0.39 & -20.2 $\pm$ 0.18 \\
5889$^{b}$ & 3.4791 & -30.3149 & 9.05 & 0.63 $\pm$ 0.12 & -19.4 $\pm$ 0.11 & 0.39 $\pm$ 0.11 & -19.0 $\pm$ 0.17 & 0.52 $\pm$ 0.13$^{a}$ & -19.1 $\pm$ 0.34 & 0.45 $\pm$ 0.09 & -19.0 $\pm$ 0.49 & 0.61 $\pm$ 0.34$^{a}$ & -18.6 $\pm$ 0.61 & -19.7 $\pm$ 0.21\\

\hline
\multicolumn{5}{l}{a.the size measurement is not confident.}\\
\multicolumn{10}{l}{b.the sources from 560 - 5889 are corresponding to GHZ1 - GHZ6 in \citetalias[][]{Castellano2022}.} \\
\multicolumn{10}{l}{c.$M_\text{UV}$, rest-frame UV absolute magnitude, calculated in \citetalias[][]{Leethochawalit2022} and \citetalias[][]{Castellano2022}.}\\ \\ \\ \\ \\ \\ \\ \\ \\ \\ \\ \\ \\ \\ \\ \\ \\ \\ \\ \\
\end{tabular}

\end{sidewaystable*}

\section{Size-luminosity distribution}\label{sec:size-lum}

\begin{figure*} 
\centering

\includegraphics[width=2\columnwidth]{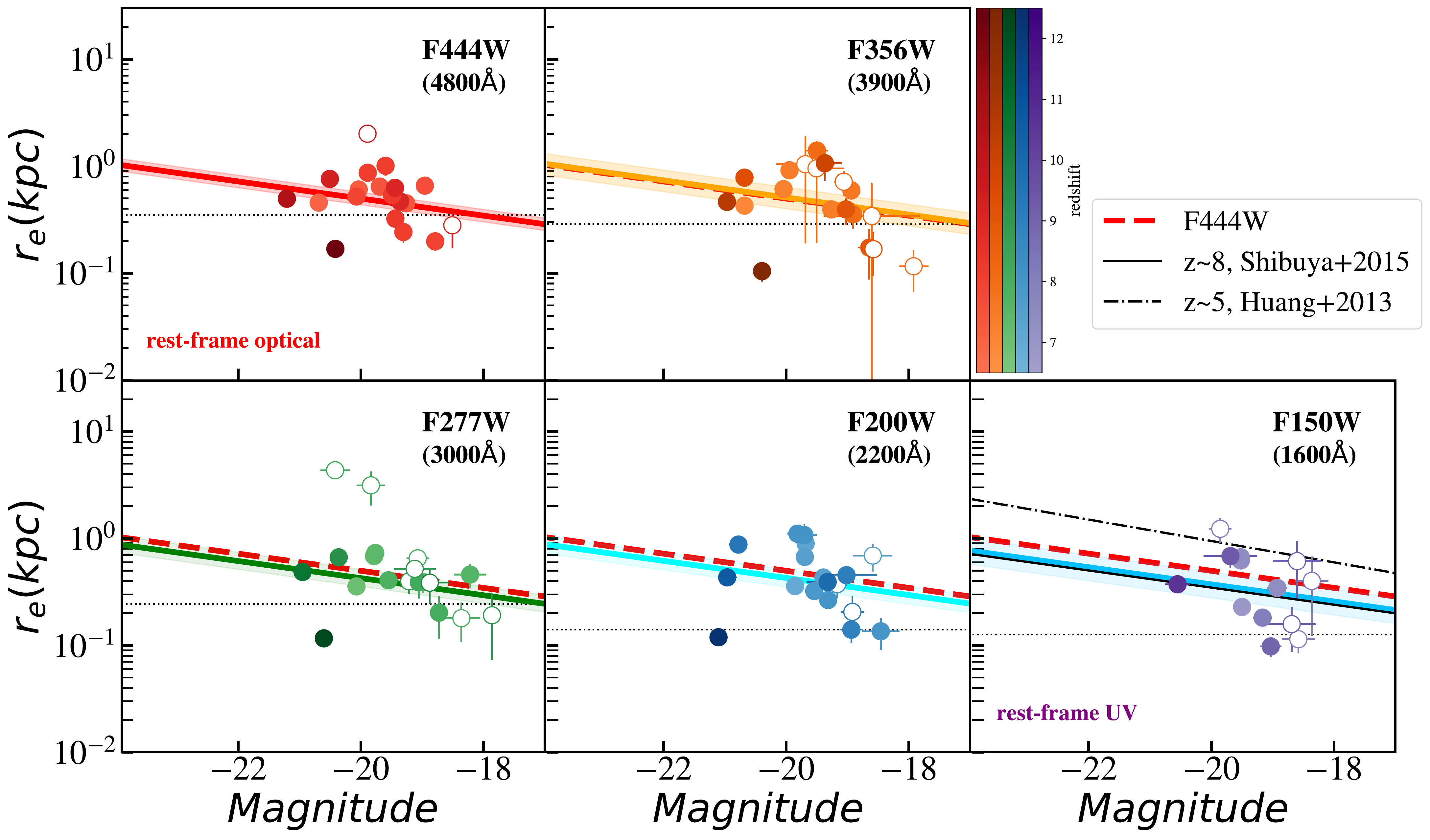}
    \caption{
    Size-luminosity distribution of galaxies observed in 5 NIRCam bands from F444W to F150W band, corresponding to rest-frame optical ($\sim4800\,{\rm \AA}$) to UV  ($\sim1600\,{\rm \AA}$).
    The x-axes represent the absolute magnitude measured in each band, and the y-axes represent the effective radius measured by \texttt{Galight} using the S\'ersic model.
    The solid points, solid lines, and shadow region in each panel represent respectively the data, the best fit of the size-luminosity relation, and $1\,\sigma$ uncertainty range of the derived relation. 
    The color transparency of the data points indicates the redshift.
    The empty circles show the galaxies with less secure size determination, displayed here for completeness but not used for fitting. The red dash lines represent the best fit obtained in the F444W band. 
    The black solid and dash-dot lines in the F150W panel show the relation derived from HST data, \citet[][$z\sim8$]{Shibuya2015} and \citet[][$z\sim5$]{Huang2013} at the similar rest-frame wavelength, respectively.
    The horizontal black dotted line in each panel indicates the PSF size (FWHM/2) for reference.
    } 
\label{fig:size-luminosity-7bands}

\end{figure*}

In Figure~\ref{fig:size-luminosity-7bands}, we present the size-luminosity distribution of galaxies at $z>7$ from the rest-frame optical ($\sim4800$\,\AA) to UV ($\sim1600$\,\AA) bands.
The source luminosity is calculated from the best-fit magnitude of our S\'ersic fits.

In this Section, we apply an analytic method to derive the size-luminosity relation and report the best-fits results. Given the relatively small sample size at $z>9$ in this first dataset, we do not consider potential evolution of the intercept with redshift.

\subsection{Size-Luminosity relation in analytical form}\label{subsec:analytic}

In order to carry out our fit,
we assume that the size distribution at a given luminosity obeys a log-normal function and that the size-luminosity relation can be described by a power-law \citep{Shen2003,Holwerda2015}. 

In this way, the probability density function (PDF) of the size-luminosity pair is expressed as,
\begin{equation}
P(r_{e},L; r_{0}, \sigma, \beta)=\frac{1}{r_e\sigma\sqrt{2\pi}} \exp \left(-\frac{({\rm ln}r_e-{\rm ln}\overline{r_e})^2}{2\sigma^2}\right)\\
\end{equation}
and 
\begin{equation}
\overline{r_e}=r_0\left (\frac{L}{L_0} \right)^\beta
\end{equation}
where r$_0$, $\sigma$, $\beta$ and $L_0$ are the median radius at $L_0$, the log-normal dispersion of $\rm{ln}r_{e}$, the slope of the size-luminosity relation, and the characteristic luminosity (corresponding to $M=-21.0$), respectively.

Given the relatively small sample size and narrow brightness range, 
for our baseline analysis we opt to fix the slope.
Since many studies have found consistent UV slope for entire redshift range \citep{Huang2013,Ono2013,Shibuya2015, Holwerda2015}, we fix the slope $\beta=0.20$ as reported by \citet{Shibuya2015} at $z\sim8$ from HST data.
We then use a standard Bayesian approach and run a Markov chain Monte Carlo (MCMC) process to derive the posterior distribution of parameters r$_0$, and $\sigma$.

\subsection{The size-luminosity relation as a function of wavelength}\label{subsec:size-lum-results}

In each panel of Figure~\ref{fig:size-luminosity-7bands}, colored points and lines represent the data and fitting results in the corresponding band. We performed size measurements on 19, 19, 19, 18, and 13 galaxies in the F444W,
F356W, F277W, F200W, and F150W band, respectively.
The failures in F200W and F150W bands are due to a combination of factors: galaxies are out of the detector, have too low SNR (i.e.,$<1.5$), or have an extremely bright contaminant nearby. Some galaxies are not confidently measured, as judged by significant residuals to the S\'ersic model fit. Those sources are identified by empty circles, They are shown for completeness and are excluded from the size-luminosity relation fit.
Some galaxies are extremely small, i.e., smaller than the PSF size.
We performed mock tests using a S\'ersic profile to model the galaxies, and it turns out that the radius can be correctly recovered down to half of the PSF size, i.e.,  $\sim30$ mas at the F444W band.

We fit our sample galaxies in the size-luminosity plane of each filter in Figure~\ref{fig:size-luminosity-7bands}. The best-fit parameters are summarized in Table~\ref{tab:best-fits}. We re-iterate that for this initial analysis, we have limited ourselves to relatively bright galaxies $\sim m_{F444W}<28$, and have neglected the effects of potential incompleteness bias and lensing magnification. We will revisit these issues in future work, after the completion of the campaign.

The red dash lines indicate the baseline measured in F444W, 
for comparison of galaxy size at different rest-frame wavelengths. The median size $r_0$ is $\sim 450-600$ pc, and slightly decreases from $600$ pc in rest-frame optical to $450$ pc in rest-frame UV. The scatter, $\sigma\sim0.7$, does not vary significantly across the filters.

In one of the panels (F150W), we show the best-fit relation of $z\sim8$ Lyman-break galaxies from \citet[][]{Shibuya2015}, measured based on HST images,
which is consistent with our results.
As a comparison to lower redshift, we also show the best-fit relation of $z\sim5$ Lyman-break galaxies from \citet{Huang2013}.

\begin{figure} 
\centering
\includegraphics[width=0.65\columnwidth]{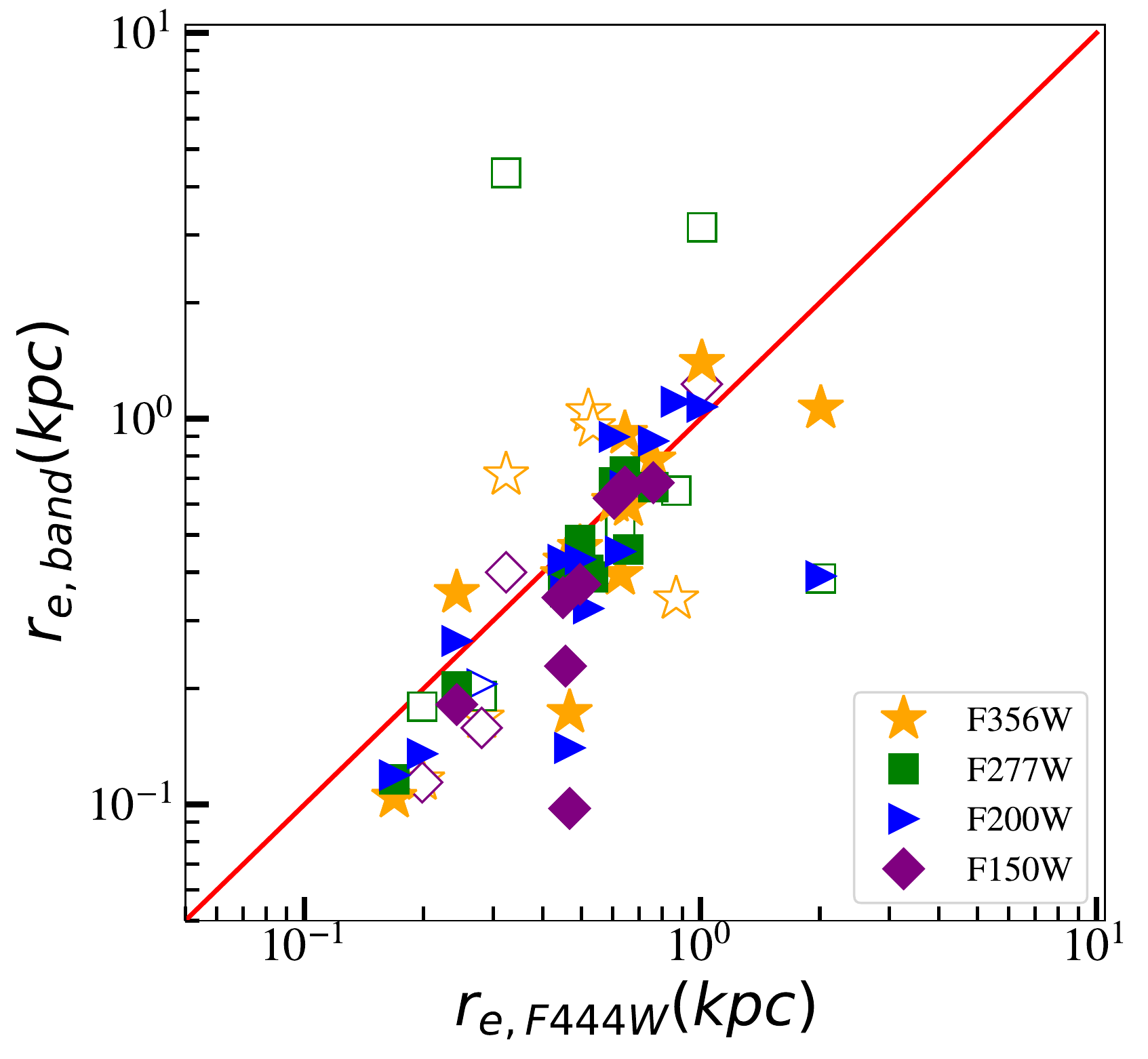}
\includegraphics[width=0.33\columnwidth]{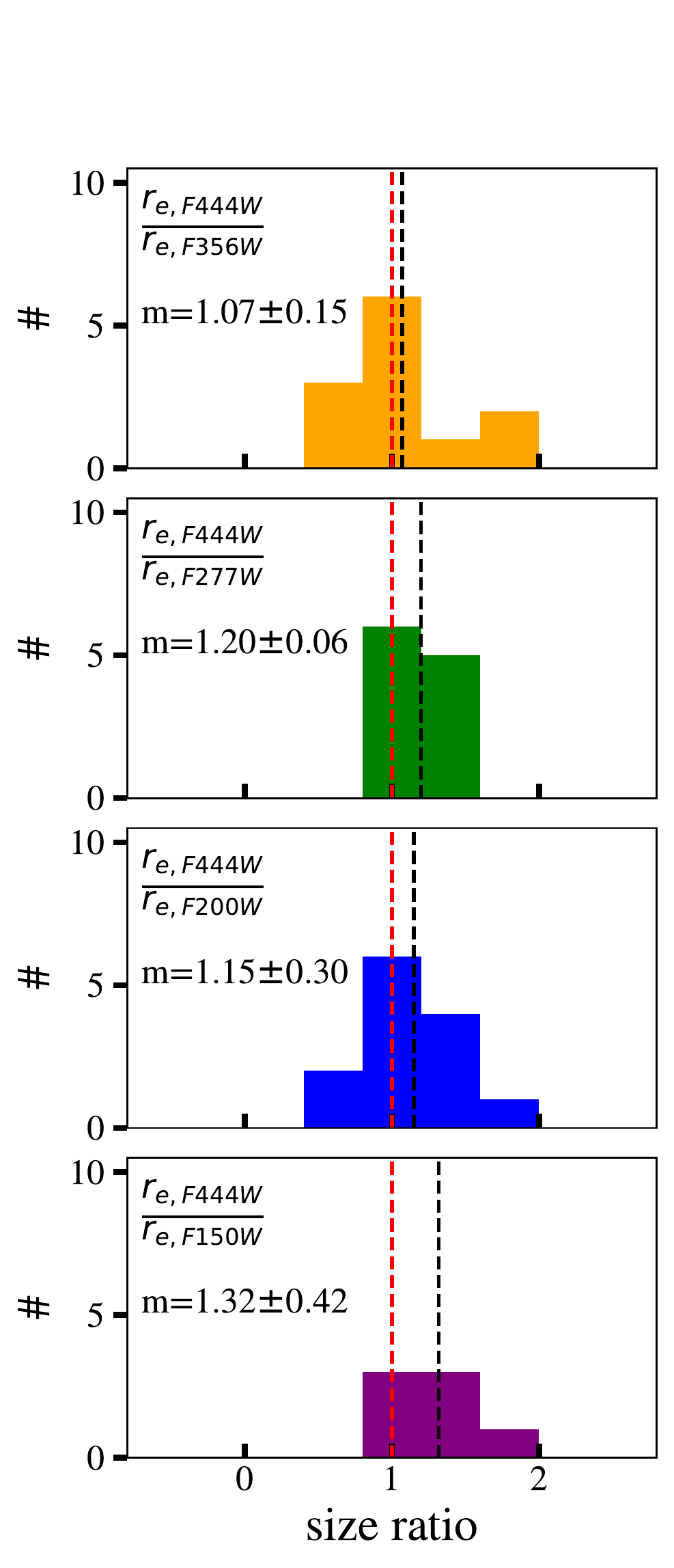}
\caption{
Size comparison among five bands.
(Left) Rest-frame optical (F444W) band size compared to those of other 4 bluer bands, i.e., F356W (orange stars), F277W (green squares), F200W (blue triangles), and F150W (purple diamonds). 
The empty symbols represent those with less secure size determination, which are not used for calculating size ratio.
(Right) Histograms of size ratio distribution between F444W and each of the 4 bluer bands.
In each panel, we indicate the median value of the distribution by the black dash line and reference (size ratio = 1) by the red dash line.}\label{fig:size-comp}
\end{figure}

Lastly, we compare the sizes obtained from the F444W band to those obtained with the other four bluer bands on an individual basis in Figure~\ref{fig:size-comp}. 
In the left panel, we see that the average and median galaxy sizes becomes smaller at shorter wavelengths, with F150W the most significantly different one. 
In the right panel, we demonstrate the size ratio distribution and find the median values are 
$1.07\pm0.15$, $1.20\pm0.06$, $1.15\pm0.30$ and $1.32\pm0.42$ for
$r_{e,\text{F444W}}$/$r_{e,\text{F356W}}$, 
$r_{e,\text{F444W}}$/$r_{e,\text{F277W}}$,
$r_{e,\text{F444W}}$/$r_{e,\text{F200W}}$,
and 
$r_{e,\text{F444W}}$/$r_{e,\text{F150W}}$,
respectively.

\begin{table}
\centering
\small
\caption{Best-fit parameters of the size-luminosity relation from the rest-frame optical to UV at $z>7$. }
\begin{tabular}{ l c c c c }
\hline
\hline
Filter & $\lambda$ & $\log_{10}r_0$/kpc & $\sigma$ \\
\hline
F444W & 4800${\rm \AA}$ & -0.22 $\pm$ 0.06 & 0.52 $\pm$0.10  \\
F356W & 3900${\rm \AA}$ & -0.21 $\pm$ 0.10 & 0.78 $\pm$0.18  \\
F277W & 3000${\rm \AA}$ & -0.29 $\pm$ 0.08 & 0.63 $\pm$0.16  \\
F200W & 2200${\rm \AA}$ & -0.29 $\pm$ 0.08 & 0.76 $\pm$0.15  \\
F150W & 1600${\rm \AA}$ & -0.35 $\pm$ 0.12 & 0.76 $\pm$0.23  \\
\hline
\end{tabular}
\label{tab:best-fits}
\end{table}

\section{Discussion \& Conclusions}\label{sec:dis-con}

In the spirit of a preliminary exploration, we re-fit the size-luminosity relation allowing the slope to vary. These results should be taken with a grain of salt considering the small sample size.
We find that the slopes in F444W, F356W, F277W, F200W and F150W are 
$0.19\pm0.09$ , $0.25\pm 0.13$ , $0.17\pm 0.10$, $0.30\pm0.13$ and $0.53 \pm 0.15$, respectively.
The slopes in F444W, F356W F277W, and F200W bands are in agreement with the slope assumed in the analysis above, $\beta\sim~0.2$.
However, the slope in F150W (rest UV) is marginally steeper than that, and rather consistent with the results reported from several studies that analysed lensed
galaxies in the Hubble Frontier Fields \citep[][]{Kawamata2018, Bouwens2022, Yang2022}.
We note that the tentative steepening is driven by the smallest and faintest galaxies which might not be resolved by HST without lensing.
If confirmed by larger samples, it would imply that the UV size-luminosity might be a broken power-law, i.e., a flatter slope at the brighter end and a steeper slope at the fainter end as suggested by \citet[][]{Bouwens2022}.

We have shown that the median size $r_0$ decrease slightly from the optical to the far UV,  
$600\pm80$ pc, to $450\pm130$ pc
at $M=-21$.  
However, we stress that the effect is small and less than we would have expected if the UV emission were confined in a small star forming region surrounded by a larger envelope of older stars. 

The moderate size variation as a function of wavelength is consistent with the analysis of other morphological parameters, i.e. the Gini structural parameter, M$_{20}$, concentration, asymmetry and smoothness, reported in the companion paper by \citet[][]{Treu2022b}.
We refer the reader to \citet{Treu2022b} for more discussion of the physical interpretation of our findings. We just caution the reader that our results apply to galaxies selected primarily through the Lyman Break technique. Heavily dust obscured or fully quiescent galaxies (if they exist) would not of course be described by our findings.

We also report a tentative detection of a steepening of the slope toward fainter end in the bluest UV filter. 

In the future, after the completion of the campaign,
we will revisit this work by including the entire sample, correcting the incompleteness bias, and taking the potential lensing effect into consideration.

\section*{Acknowledgments}
\begin{acknowledgements}
This work is based on observations made with the NASA/ESA/CSA James Webb Space Telescope. The data were obtained from the Mikulski Archive for Space Telescopes at the Space Telescope Science Institute, which is operated by the Association of Universities for Research in Astronomy, Inc., under NASA contract NAS 5-03127 for JWST. These observations are associated with program JWST-ERS-1324. We acknowledge financial support from NASA through grant JWST-ERS-1324.
LY acknowledges support by JSPS KAKENHI Grant Number JP 21F21325.
KG and TN acknowledge support from Australian Research Council Laureate Fellowship FL180100060.
MB acknowledges support from the Slovenian national research agency ARRS through grant N1-0238.
CM acknowledges support by the VILLUM FONDEN under grant 37459. The Cosmic Dawn Center (DAWN) is funded by the Danish National Research Foundation under grant DNRF140.
\end{acknowledgements}

\bibliography{sample631}{}

\begin{thebibliography}{}
\expandafter\ifx\csname natexlab\endcsname\relax\def\natexlab#1{#1}\fi
\providecommand{\url}[1]{\href{#1}{#1}}
\providecommand{\dodoi}[1]{doi:~\href{http://doi.org/#1}{\nolinkurl{#1}}}
\providecommand{\doeprint}[1]{\href{http://ascl.net/#1}{\nolinkurl{http://ascl.net/#1}}}
\providecommand{\doarXiv}[1]{\href{https://arxiv.org/abs/#1}{\nolinkurl{https://arxiv.org/abs/#1}}}

\bibitem[{{Bergamini} {et~al.}(2022){Bergamini}, {Acebron}, {Grillo}, {Rosati},
  {Caminha}, {Mercurio}, {Vanzella}, {Angora}, {Brammer}, {Meneghetti}, \&
  {Nonino}}]{Bergamini2022}
{Bergamini}, P., {Acebron}, A., {Grillo}, C., {et~al.} 2022, arXiv:2207.09416,
  arXiv:2207.09416.
\newblock \doarXiv{2207.09416}

\bibitem[{{Birrer} \& {Amara}(2018)}]{Birrer&Amara2018}
{Birrer}, S., \& {Amara}, A. 2018, Physics of the Dark Universe, 22, 189,
  \dodoi{10.1016/j.dark.2018.11.002}

\bibitem[{{Birrer} {et~al.}(2015){Birrer}, {Amara}, \&
  {Refregier}}]{Birrer2015}
{Birrer}, S., {Amara}, A., \& {Refregier}, A. 2015, \apj, 813, 102,
  \dodoi{10.1088/0004-637X/813/2/102}

\bibitem[{{Birrer} {et~al.}(2021){Birrer}, {Shajib}, {Gilman}, {Galan},
  {Aalbers}, {Millon}, {Morgan}, {Pagano}, {Park}, {Teodori}, {Tessore},
  {Ueland}, {Van de Vyvere}, {Wagner-Carena}, {Wempe}, {Yang}, {Ding},
  {Schmidt}, {Sluse}, {Zhang}, \& {Amara}}]{Birrer2021}
{Birrer}, S., {Shajib}, A., {Gilman}, D., {et~al.} 2021, The Journal of Open
  Source Software, 6, 3283, \dodoi{10.21105/joss.03283}

\bibitem[{{Bouwens} {et~al.}(2022){Bouwens}, {Illingworth}, {van Dokkum},
  {Oesch}, {Stefanon}, \& {Ribeiro}}]{Bouwens2022}
{Bouwens}, R.~J., {Illingworth}, G.~D., {van Dokkum}, P.~G., {et~al.} 2022,
  \apj, 927, 81, \dodoi{10.3847/1538-4357/ac4791}

\bibitem[{{Castellano} {et~al.}(2022){Castellano}, {Fontana}, {Treu},
  {Santini}, {Merlin}, {Leethochawalit}, {Trenti}, {Mestric}, {Vanzella},
  {Bonchi}, {Belfiori}, {Nonino}, {Paris}, {Polenta}, {Roberts-Borsani},
  {Boyett}, {Calabro}, {Glazebrook}, {Grillo}, {Mascia}, {Mason}, {Mercurio},
  {Morishita}, {Nanayakkara}, {Pentericci}, {Rosati}, {Vulcani}, {Wang}, \&
  {Yang}}]{Castellano2022}
{Castellano}, M., {Fontana}, A., {Treu}, T., {et~al.} 2022, arXiv e-prints,
  arXiv:2207.09436.
\newblock \doarXiv{2207.09436}

\bibitem[{{Conselice}(2014)}]{Conselice2014}
{Conselice}, C.~J. 2014, \araa, 52, 291,
  \dodoi{10.1146/annurev-astro-081913-040037}

\bibitem[{{Ding} {et~al.}(2020){Ding}, {Silverman}, {Treu}, {Schulze},
  {Schramm}, {Birrer}, {Park}, {Jahnke}, {Bennert}, {Kartaltepe}, {Koekemoer},
  {Malkan}, \& {Sanders}}]{Ding2020}
{Ding}, X., {Silverman}, J., {Treu}, T., {et~al.} 2020, \apj, 888, 37,
  \dodoi{10.3847/1538-4357/ab5b90}

\bibitem[{{Doyon} {et~al.}(2012){Doyon}, {Hutchings}, {Beaulieu}, {Albert},
  {Lafreni{\`e}re}, {Willott}, {Touahri}, {Rowlands}, {Maszkiewicz},
  {Fullerton}, {Volk}, {Martel}, {Chayer}, {Sivaramakrishnan}, {Abraham},
  {Ferrarese}, {Jayawardhana}, {Johnstone}, {Meyer}, {Pipher}, \&
  {Sawicki}}]{NIRISS}
{Doyon}, R., {Hutchings}, J.~B., {Beaulieu}, M., {et~al.} 2012, in Society of
  Photo-Optical Instrumentation Engineers (SPIE) Conference Series, Vol. 8442,
  Space Telescopes and Instrumentation 2012: Optical, Infrared, and Millimeter
  Wave, ed. M.~C. {Clampin}, G.~G. {Fazio}, H.~A. {MacEwen}, \& J.~{Oschmann},
  Jacobus~M., 84422R, \dodoi{10.1117/12.926578}

\bibitem[{{Ferruit} {et~al.}(2022){Ferruit}, {Jakobsen}, {Giardino}, {Rawle},
  {Alves de Oliveira}, {Arribas}, {Beck}, {Birkmann}, {B{\"o}ker}, {Bunker},
  {Chariot}, {de Marchi}, {Franx}, {Henry}, {Karakla}, {Kassin}, {Kumari},
  {L{\'o}pez-Caniego}, {L{\"u}tzgendorf}, {Maiolino}, {Manjavacas}, {Marston},
  {Moseley}, {Muzerolle}, {Pirzkal}, {Rauscher}, {Rix}, {Sabbi}, {Sirianni},
  {te Plate}, {Valenti}, {Willott}, \& {Zeidler}}]{NIRSPEC2}
{Ferruit}, P., {Jakobsen}, P., {Giardino}, G., {et~al.} 2022, \aap, 661, A81,
  \dodoi{10.1051/0004-6361/202142673}

\bibitem[{{Fukugita} {et~al.}(1996){Fukugita}, {Ichikawa}, {Gunn}, {Doi},
  {Shimasaku}, \& {Schneider}}]{fukugita96}
{Fukugita}, M., {Ichikawa}, T., {Gunn}, J.~E., {et~al.} 1996, \aj, 111, 1748,
  \dodoi{10.1086/117915}

\bibitem[{{Grazian} {et~al.}(2012){Grazian}, {Castellano}, {Fontana},
  {Pentericci}, {Dunlop}, {McLure}, {Koekemoer}, {Dickinson}, {Faber},
  {Ferguson}, {Galametz}, {Giavalisco}, {Grogin}, {Hathi}, {Kocevski}, {Lai},
  {Newman}, \& {Vanzella}}]{Grazian2012}
{Grazian}, A., {Castellano}, M., {Fontana}, A., {et~al.} 2012, \aap, 547, A51,
  \dodoi{10.1051/0004-6361/201219669}

\bibitem[{{Holwerda} {et~al.}(2015){Holwerda}, {Bouwens}, {Oesch}, {Smit},
  {Illingworth}, \& {Labbe}}]{Holwerda2015}
{Holwerda}, B.~W., {Bouwens}, R., {Oesch}, P., {et~al.} 2015, \apj, 808, 6,
  \dodoi{10.1088/0004-637X/808/1/6}

\bibitem[{{Huang} {et~al.}(2013){Huang}, {Ferguson}, {Ravindranath}, \&
  {Su}}]{Huang2013}
{Huang}, K.-H., {Ferguson}, H.~C., {Ravindranath}, S., \& {Su}, J. 2013, \apj,
  765, 68, \dodoi{10.1088/0004-637X/765/1/68}

\bibitem[{{Jakobsen} {et~al.}(2022){Jakobsen}, {Ferruit}, {Alves de Oliveira},
  {Arribas}, {Bagnasco}, {Barho}, {Beck}, {Birkmann}, {B{\"o}ker}, {Bunker},
  {Charlot}, {de Jong}, {de Marchi}, {Ehrenwinkler}, {Falcolini}, {Fels},
  {Franx}, {Franz}, {Funke}, {Giardino}, {Gnata}, {Holota}, {Honnen}, {Jensen},
  {Jentsch}, {Johnson}, {Jollet}, {Karl}, {Kling}, {K{\"o}hler}, {Kolm},
  {Kumari}, {Lander}, {Lemke}, {L{\'o}pez-Caniego}, {L{\"u}tzgendorf},
  {Maiolino}, {Manjavacas}, {Marston}, {Maschmann}, {Maurer}, {Messerschmidt},
  {Moseley}, {Mosner}, {Mott}, {Muzerolle}, {Pirzkal}, {Pittet}, {Plitzke},
  {Posselt}, {Rapp}, {Rauscher}, {Rawle}, {Rix}, {R{\"o}del}, {Rumler},
  {Sabbi}, {Salvignol}, {Schmid}, {Sirianni}, {Smith}, {Strada}, {te Plate},
  {Valenti}, {Wettemann}, {Wiehe}, {Wiesmayer}, {Willott}, {Wright}, {Zeidler},
  \& {Zincke}}]{NIRSPEC1}
{Jakobsen}, P., {Ferruit}, P., {Alves de Oliveira}, C., {et~al.} 2022, \aap,
  661, A80, \dodoi{10.1051/0004-6361/202142663}

\bibitem[{{Kawamata} {et~al.}(2018){Kawamata}, {Ishigaki}, {Shimasaku},
  {Oguri}, {Ouchi}, \& {Tanigawa}}]{Kawamata2018}
{Kawamata}, R., {Ishigaki}, M., {Shimasaku}, K., {et~al.} 2018, \apj, 855, 4,
  \dodoi{10.3847/1538-4357/aaa6cf}

\bibitem[{{Kawinwanichakij} {et~al.}(2021){Kawinwanichakij}, {Silverman},
  {Ding}, {George}, {Damjanov}, {Sawicki}, {Tanaka}, {Taranu}, {Birrer},
  {Huang}, {Li}, {Onodera}, {Shibuya}, \& {Yasuda}}]{Kawinwanichakij2021}
{Kawinwanichakij}, L., {Silverman}, J.~D., {Ding}, X., {et~al.} 2021, \apj,
  921, 38, \dodoi{10.3847/1538-4357/ac1f21}

\bibitem[{{Leethochawalit} {et~al.}(2022){Leethochawalit}, {Trenti}, {Santini},
  {Yang}, {Merlin}, {Castellano}, {Fontana}, {Treu}, {Mason}, {Glazebrook},
  {Jones}, {Vulcani}, {Nanayakkara}, {Marchesini}, {Mascia}, {Morishita},
  {Roberts-Borsani}, {Bonchi}, {Paris}, {Boyett}, {Strait}, {Calabro`},
  {Pentericci}, {Bradac}, {Wang}, \& {Scarlata}}]{Leethochawalit2022}
{Leethochawalit}, N., {Trenti}, M., {Santini}, P., {et~al.} 2022, arXiv
  e-prints, arXiv:2207.11135.
\newblock \doarXiv{2207.11135}

\bibitem[{{Ma} {et~al.}(2018){Ma}, {Hopkins}, {Boylan-Kolchin},
  {Faucher-Gigu{\`e}re}, {Quataert}, {Feldmann}, {Garrison-Kimmel}, {Hayward},
  {Kere{\v{s}}}, \& {Wetzel}}]{Ma2018}
{Ma}, X., {Hopkins}, P.~F., {Boylan-Kolchin}, M., {et~al.} 2018, \mnras, 477,
  219, \dodoi{10.1093/mnras/sty684}

\bibitem[{{Marshall} {et~al.}(2022){Marshall}, {Wilkins}, {Di Matteo}, {Roper},
  {Vijayan}, {Ni}, {Feng}, \& {Croft}}]{Marshall2022}
{Marshall}, M.~A., {Wilkins}, S., {Di Matteo}, T., {et~al.} 2022, \mnras, 511,
  5475, \dodoi{10.1093/mnras/stac380}

\bibitem[{{Medezinski} {et~al.}(2016){Medezinski}, {Umetsu}, {Okabe}, {Nonino},
  {Molnar}, {Massey}, {Dupke}, \& {Merten}}]{Medezinski2016}
{Medezinski}, E., {Umetsu}, K., {Okabe}, N., {et~al.} 2016, \apj, 817, 24,
  \dodoi{10.3847/0004-637X/817/1/24}

\bibitem[{{Merlin} {et~al.}(2022){Merlin}, {Bonchi}, {Paris}, {Belfiori},
  {Fontana}, {Castellano}, {Nonino}, {Polenta}, {Santini}, {Yang},
  {Glazebrook}, {Treu}, {Roberts-Borsani}, {Trenti}, {Birrer}, {Brammer},
  {Grillo}, {Calabr{\`o}}, {Marchesini}, {Mason}, {Mercurio}, {Morishita},
  {Strait}, {Boyett}, {Leethochawalit}, {Nanayakkara}, {Vulcani}, {Bradac}, \&
  {Wang}}]{Merlin2022}
{Merlin}, E., {Bonchi}, A., {Paris}, D., {et~al.} 2022, arXiv e-prints,
  arXiv:2207.11701.
\newblock \doarXiv{2207.11701}

\bibitem[{{Morishita} {et~al.}(2014){Morishita}, {Ichikawa}, \&
  {Kajisawa}}]{Morishita2014}
{Morishita}, T., {Ichikawa}, T., \& {Kajisawa}, M. 2014, \apj, 785, 18,
  \dodoi{10.1088/0004-637X/785/1/18}

\bibitem[{{Oke} \& {Gunn}(1983)}]{oke83}
{Oke}, J.~B., \& {Gunn}, J.~E. 1983, \apj, 266, 713, \dodoi{10.1086/160817}

\bibitem[{{Ono} {et~al.}(2013){Ono}, {Ouchi}, {Curtis-Lake}, {Schenker},
  {Ellis}, {McLure}, {Dunlop}, {Robertson}, {Koekemoer}, {Bowler}, {Rogers},
  {Schneider}, {Charlot}, {Stark}, {Shimasaku}, {Furlanetto}, \&
  {Cirasuolo}}]{Ono2013}
{Ono}, Y., {Ouchi}, M., {Curtis-Lake}, E., {et~al.} 2013, \apj, 777, 155,
  \dodoi{10.1088/0004-637X/777/2/155}

\bibitem[{{Peng} {et~al.}(2002){Peng}, {Ho}, {Impey}, \& {Rix}}]{Peng2002}
{Peng}, C.~Y., {Ho}, L.~C., {Impey}, C.~D., \& {Rix}, H.-W. 2002, \aj, 124,
  266, \dodoi{10.1086/340952}

\bibitem[{{Rieke} {et~al.}(2005){Rieke}, {Kelly}, \& {Horner}}]{NIRCAM}
{Rieke}, M.~J., {Kelly}, D., \& {Horner}, S. 2005, in Society of Photo-Optical
  Instrumentation Engineers (SPIE) Conference Series, Vol. 5904, Cryogenic
  Optical Systems and Instruments XI, ed. J.~B. {Heaney} \& L.~G. {Burriesci},
  1--8, \dodoi{10.1117/12.615554}

\bibitem[{{Roberts-Borsani} {et~al.}(2022){Roberts-Borsani}, {Morishita},
  {Treu}, {Brammer}, {Strait}, {Wang}, {Bradac}, {Acebron}, {Bergamini},
  {Boyett}, {Calabr{\'o}}, {Castellano}, {Fontana}, {Glazebrook}, {Grillo},
  {Henry}, {Jones}, {Malkan}, {Marchesini}, {Mascia}, {Mason}, {Mercurio},
  {Merlin}, {Nanayakkara}, {Pentericci}, {Rosati}, {Santini}, {Scarlata},
  {Trenti}, {Vanzella}, {Vulcani}, \& {Willott}}]{Roberts-Borsani2022}
{Roberts-Borsani}, G., {Morishita}, T., {Treu}, T., {et~al.} 2022, arXiv
  e-prints, arXiv:2207.11387.
\newblock \doarXiv{2207.11387}

\bibitem[{{Roper} {et~al.}(2022){Roper}, {Lovell}, {Vijayan}, {Marshall},
  {Irodotou}, {Kuusisto}, {Thomas}, \& {Wilkins}}]{Roper2022}
{Roper}, W.~J., {Lovell}, C.~C., {Vijayan}, A.~P., {et~al.} 2022, \mnras,
  \dodoi{10.1093/mnras/stac1368}

\bibitem[{{Sersic}(1968)}]{Sersic1968}
{Sersic}, J.~L. 1968, {Atlas de Galaxias Australes}

\bibitem[{{Shen} {et~al.}(2003){Shen}, {Mo}, {White}, {Blanton}, {Kauffmann},
  {Voges}, {Brinkmann}, \& {Csabai}}]{Shen2003}
{Shen}, S., {Mo}, H.~J., {White}, S. D.~M., {et~al.} 2003, \mnras, 343, 978,
  \dodoi{10.1046/j.1365-8711.2003.06740.x}

\bibitem[{{Shibuya} {et~al.}(2015){Shibuya}, {Ouchi}, \&
  {Harikane}}]{Shibuya2015}
{Shibuya}, T., {Ouchi}, M., \& {Harikane}, Y. 2015, \apjs, 219, 15,
  \dodoi{10.1088/0067-0049/219/2/15}

\bibitem[{{Treu} {et~al.}(2022{\natexlab{a}}){Treu}, {Roberts-Borsani},
  {Bradac}, {Brammer}, {Fontana}, {Henry}, {Mason}, {Morishita}, {Pentericci},
  {Wang}, {Acebron}, {Bagley}, {Bergamini}, {Belfiori}, {Bonchi}, {Boyett},
  {Boutsia}, {Calabro}, {Caminha}, {Castellano}, {Dressler}, {Glazebrook},
  {Grillo}, {Jacobs}, {Jones}, {Kelly}, {Leethochawalit}, {Malkan},
  {Marchesini}, {Mascia}, {Mercurio}, {Merlin}, {Nanayakkara}, {Paris},
  {Poggianti}, {Rosati}, {Santini}, {Scarlata}, {Shipley}, {Strait}, {Trenti},
  {Tubthong}, {Vanzella}, {Vulcani}, \& {Yang}}]{Treu2022}
{Treu}, T., {Roberts-Borsani}, G., {Bradac}, M., {et~al.} 2022{\natexlab{a}},
  arXiv e-prints, arXiv:2206.07978.
\newblock \doarXiv{2206.07978}

\bibitem[{{Treu} {et~al.}(2022{\natexlab{b}}){Treu}, {Calabro}, {Castellano},
  {Leethochawalit}, {Merlin}, {Fontana}, {Yang}, {Morishita}, {Trenti},
  {Dressler}, {Mason}, {Paris}, {Pentericci}, {Roberts-Borsani}, {Vulcani},
  {Boyett}, {Bradac}, {Glazebrook}, {Jones}, {Marchesini}, {Mascia},
  {Nanayakkara}, {Santini}, {Strait}, {Vanzella}, \& {Wang}}]{Treu2022b}
{Treu}, T., {Calabro}, A., {Castellano}, M., {et~al.} 2022{\natexlab{b}}, arXiv
  e-prints, arXiv:2207.13527.
\newblock \doarXiv{2207.13527}

\bibitem[{{Wu} {et~al.}(2020){Wu}, {Dav{\'e}}, {Tacchella}, \& {Lotz}}]{Wu2020}
{Wu}, X., {Dav{\'e}}, R., {Tacchella}, S., \& {Lotz}, J. 2020, \mnras, 494,
  5636, \dodoi{10.1093/mnras/staa1044}

\bibitem[{{Yang} {et~al.}(2021){Yang}, {Roberts-Borsani}, {Treu}, {Birrer},
  {Morishita}, \& {Brada{\v{c}}}}]{Yang2021}
{Yang}, L., {Roberts-Borsani}, G., {Treu}, T., {et~al.} 2021, \mnras, 501,
  1028, \dodoi{10.1093/mnras/staa3713}

\bibitem[{{Yang} {et~al.}(2022){Yang}, {Leethochawalit}, {Treu},
  {Roberts-Borsani}, {Brada{\v{c}}}, {Birrer}, {Castellano}, {Merlin},
  {Fontana}, {Amorin}, \& {Trenti}}]{Yang2022}
{Yang}, L., {Leethochawalit}, N., {Treu}, T., {et~al.} 2022, \mnras, 514, 1148,
  \dodoi{10.1093/mnras/stac1236}

\end{thebibliography}
\bibliographystyle{aasjournal}

\end{document}